\documentclass[journal]{IEEEtran}
\ifCLASSINFOpdf
\else
\fi

\usepackage{tabularx}
\usepackage{cite}
\usepackage{amssymb}
\usepackage{pifont}
\usepackage{adjustbox}
\usepackage{amsmath}
\usepackage{amsfonts}
\usepackage{graphicx}
\usepackage{textcomp}
\usepackage{xcolor}
\usepackage{multirow}
\usepackage{soul}
\usepackage{tikz}
\usepackage{wrapfig}
\usepackage{algorithm}
\usepackage{algpseudocode}
\usepackage{colortbl}
\usepackage{bm}
\usepackage{float}
\usepackage{filecontents}
\usepackage{ragged2e}
\usepackage[dvipsnames]{xcolor}
\usepackage{enumitem}
\usepackage{subcaption} 
\usepackage{arydshln}
\usepackage{tabularx}
\usepackage{adjustbox}
\usepackage{booktabs}    
\usepackage{makecell}

\newcommand{\mytbo}{\cellcolor{orange!50}}
\newcommand{\mytbg}{\cellcolor{green!15}}

\setlist{nolistsep}
\usepackage{xcolor} 

\definecolor{darkgreen}{RGB}{0, 100, 0} 

\newcommand{\cmmnt}[1]{}  

\def\BibTeX{{\rm B\kern-.05em{\sc i\kern-.025em b}\kern-.08em
    T\kern-.1667em\lower.7ex\hbox{E}\kern-.125emX}}

\newcommand\encircle[1]{%
\tikz[baseline=(X.base)] 
  \node (X) [draw, scale=0.75, shape=circle, inner sep=0, fill=black, text=white, minimum size=0em] {\strut #1};}

\usepackage{todonotes}

\usepackage{pifont}

\definecolor{mygreen}{RGB}{9, 181, 55}

\begin{document}


\title{GLANCE: \underline{G}aze-\underline{L}ed \underline{A}ttention \underline{N}etwork for \underline{C}ompressed \underline{E}dge-inference}
%
%
%

\author{
Neeraj Solanki,~\IEEEmembership{Student Member,~IEEE,}
Hong~Ding, 
Sepehr~Tabrizchi,~\IEEEmembership{Graduate Student Member,~IEEE,}
        Ali Shafiee Sarvestani,~\IEEEmembership{Graduate Student Member,~IEEE,}
        Shaahin Angizi,~\IEEEmembership{Senior Member,~IEEE,}
        David~Z. Pan,~\IEEEmembership{Fellow,~IEEE,}
        and
        Arman~Roohi,~\IEEEmembership{Senior Member,~IEEE}\vspace{-2em}
\IEEEcompsocitemizethanks{\IEEEcompsocthanksitem Neeraj Solanki, Hong Ding, Sepehr Tabrizchi, Ali Shafiee Sarvestani, and Arman Roohi are with the Department of Electrical and Computer Engineering, University of Illinois Chicago, Chicago IL, USA. E-mail: aroohi@uic.edu.}
\thanks{S. Angizi is with the Department of Electrical and Computer Engineering, New Jersey Institute of Technology (NJIT), Newark, NJ, USA e-mail: (shaahin.angizi@njit.edu).}
\IEEEcompsocitemizethanks{\IEEEcompsocthanksitem D. Z. Pan is with the Department of Electrical and Computer Engineering, The University of Texas at Austin, Austin, TX 78712, USA. E-mail: dpan@mail.utexas.edu.}
\thanks{This work is supported in part by the National Science Foundation (NSF) under grant numbers 2504839, 2448133, 2303116, and 2447566.}}

%
%

\markboth{IEEE Transactions on Very Large Scale Integration (VLSI) Systems,~Vol.~ , No.~ , March~2026}%
{N. Solanki \MakeLowercase{\textit{et al.}}: Bare Demo of IEEEtran.cls for IEEE Journals}
%



\maketitle

\begin{abstract}
Real-time object detection in AR/VR systems faces critical computational constraints, requiring sub-10\,ms latency within tight power budgets. Inspired by biological foveal vision, we propose a two-stage pipeline that combines differentiable weightless neural networks for ultra-efficient gaze estimation with attention-guided region-of-interest object detection. Our approach eliminates arithmetic-intensive operations by performing gaze tracking through memory lookups rather than multiply-accumulate computations, achieving an angular error of $8.32^{\circ}$ with only 393 MACs and 2.2 KiB of memory per frame. Gaze predictions guide selective object detection on attended regions, reducing computational burden by 40-50\% and energy consumption by 65\%. Deployed on the Arduino Nano 33 BLE, our system achieves 48.1\% mAP on COCO (51.8\% on attended objects) while maintaining sub-10\,ms latency, meeting stringent AR/VR requirements by improving the communication time by $\times 177$. Compared to the global YOLOv12n baseline, which achieves 39.2\%, 63.4\%, and 83.1\% accuracy for small, MEDium, and LARGE objects, respectively, the ROI-based method yields 51.3\%, 72.1\%, and 88.1\% under the same settings. This work shows that memory-centric architectures with explicit attention modeling offer better efficiency and accuracy for resource-constrained wearable platforms than uniform processing.
\end{abstract}


%
\IEEEpeerreviewmaketitle

\section{Introduction}
%
%
%
%
\IEEEPARstart{A}{ugmented}/virtual reality (AR/VR) headsets, mobile vision apps, and wearables increasingly demand \emph{real-time} object detection under tight power and latency budgets \cite{dac1}. Contemporary devices (e.g., Meta Quest Pro, Apple Vision Pro) must concurrently perform scene understanding, object detection, gaze tracking, and high-fidelity rendering, typically within \(\leq\)tens of watts and with motion-to-photon latency below \(10\,\mathrm{ms}\) to maintain presence and avoid cybersickness~\cite{caserman2021cybersickness}. Although state-of-the-art detectors attain strong benchmark accuracy (e.g., YOLOv9 at 55.6\% mAP on COCO~\cite{wang2024yolov9}), inference often entails \(>\!10^{11}\) FLOPs per frame and substantial energy, complicating continuous operation at 60–120\,Hz on battery-powered platforms.
A central inefficiency persists: most detectors expend computation \emph{uniformly} across the full field of view. In contrast, human vision is highly selective. The fovea spans \(\sim\!2^\circ\) of visual angle and commands the majority of neural resources, while the periphery is processed at a much lower resolution~\cite{curcio1990human}, yielding on the order of \(10^2\!-\!10^3\times\) disparity in effective resource allocation between central and peripheral vision~\cite{wandell2022computational}. Computational analogs of foveation can deliver up to \(10\times\) speedups with modest accuracy loss~\cite{guenter2012foveated,meng2018kernel}, yet \emph{explicit} attention signals (e.g., gaze) remain under-utilized in practical, energy-constrained detection pipelines.

\begin{figure*}[t]
\centering
\includegraphics[width=\linewidth]{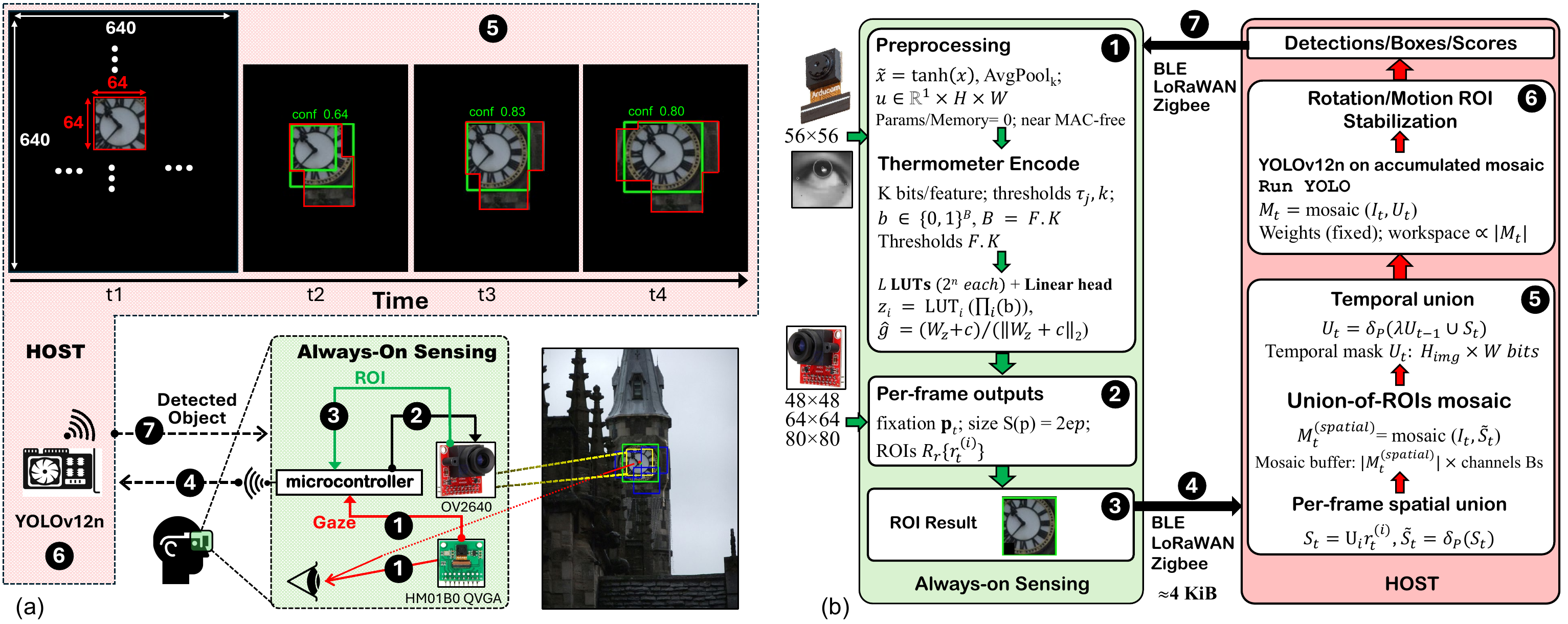}
\caption{(a) \textbf{\texttt{GLANCE}} overview: DWN-based gaze estimation drives union-of-ROIs attention; a periodic/conditional policy ships a single mosaic crop to the detector. (b) MCU/host dataflow. The host forms a per-frame spatial union to produce one union-of-ROIs mosaic and maintains a temporal state $U_t$ to trigger YOLO under a periodic/budget policy.}
\label{fig:end-to-end}
\vspace{-1em}
\end{figure*}

We address this gap with \textbf{\texttt{GLANCE}} (Fig.~\ref{fig:end-to-end}), a two-stage, attention-guided detection pipeline that pairs \emph{Differentiable Weightless Neural Networks} (DWNs) for ultra-efficient gaze estimation with \emph{selective, region-of-interest (ROI)–based} object detection. DWNs perform inference via lookup-table (LUT) operations rather than MAC-heavy layers, reducing arithmetic intensity and favoring regular, small-footprint memory access. In \textbf{\texttt{GLANCE}}, the DWN runs at frame rate to localize attended regions; the host then forms a single \emph{union-of-ROIs} mosaic $M_t$ from the current frame and feeds only this crop to the detector. 
A lightweight temporal policy maintains a $K$-frame accumulated union and \emph{invokes the detector on every frame} ($R{=}1$) using the accumulated mosaic, stabilizing recall while confining compute and memory to the union crop.
 Because the detector compute and memory scale with the mosaic area $\lvert M_t\rvert$, \textbf{\texttt{GLANCE}} concentrates resources where evidence accumulates, allowing accurate detection under tight power and memory constraints.
This work makes the following contributions:
(i) \textsl{Memory-based gaze estimation at the edge:} To our knowledge, this is the first application of weightless neural networks to real-time gaze estimation for attention-guided vision, demonstrating LUT-based inference as a viable, low-energy alternative to MAC-centric models.
(ii) \textsl{Attention-guided detection via union-of-ROIs mosaic:} We integrate explicit gaze with a memory-light ROI fusion mechanism that forms a single detector-ready mosaic, yielding a clearer efficiency–accuracy trade-off than learned-attention or saliency-only baselines.
(iii) \textsl{Temporal ROI integration policy:}  We maintain a scanpath-aware union over the last $K$ frames,
$U_t$, and perform per-frame detection on the accumulated mosaic
$M_t=\mathrm{mosaic}(I_t,U_t)$, sustaining spatial coherence and stable recall at modest compute cost.
(iv) \textsl{Rotation- and motion-aware ROI stabilization:} We reproject ROIs into a persistent $360^\circ$ map and enhance box sizes using IMU data between detector runs to maintain recall during head movements.
(v) \textsl{Resource-constrained deployment study:} On microcontroller-class platforms, \textbf{\texttt{GLANCE}} significantly reduces compute/energy and achieves lower end-to-end latency, while maintaining competitive accuracy.
By uniting explicit attention cues with memory-centric computation, \textbf{\texttt{GLANCE}} provides a practical path to \emph{attention-aware object detection} on resource-limited devices.

\section{Background}

\subsection{ Weightless Neural Networks: Memory-Centric Computation}
Weightless Neural Networks (WNNs) depart from conventional deep neural networks (DNNs) by storing knowledge directly in lookup tables (LUTs) addressed by input features, rather than in floating-point weights and MAC-intensive layers~\cite{aleksander1984wisard}. Each neuron behaves as a RAM node whose activation is obtained by reading (or updating) counters at addresses indexed by a (possibly quantized) input pattern. This memory-centric design yields three properties desirable for edge deployment: (i) \emph{zero-MAC inference} via table lookups, (ii) \emph{intrinsic parallelism} because RAM nodes are independent, and (iii) \emph{fast one-shot/online learning} through counter updates. Regular, localized memory access with small working sets maps naturally to microcontrollers (MCUs), FPGAs, and near-memory architectures.

\subsection{Classical and Differentiable Variants}
The WiSARD model is a canonical WNN~\cite{aleksander1984wisard}; each class is represented by an ensemble of RAM nodes; training increments class counters at addresses derived from the (binary) input; inference sums class-specific responses and selects the maximum. Although effective for early pattern recognition, the classical WiSARD is constrained by binary inputs and nondifferentiable operations.
Subsequent work mitigates these constraints by enabling continuous inputs and gradient-based optimization. Fuzzy, thermometer, and probabilistic encodings quantize real-valued features into ordered binary patterns compatible with LUT addressing~\cite{torres2013fuzzywisard}. Differentiable Weightless Networks (DWNs)~\cite{pmlr-v235-bacellar24a} replace static counters with trainable LUT tensors and relax discrete lookups into differentiable operators, allowing end-to-end optimization with modern regularizers. Recent systems demonstrate that weightless (or quasi-weightless) blocks can substitute dense MLPs in vision pipelines, eliminating a large fraction of multiplications while maintaining competitive accuracy and substantially reducing energy per inference~\cite{batista2023quasiweightless}.
Because WNN computation is dominated by memory access rather than arithmetic, WNNs align with in-memory/near-memory trends \cite{dac4,dac5} and the constraints of ultra-low-power embedded platforms \cite{dac2,dac3}. Studies on edge classification, tracking, and change detection report gains in latency and energy under tight power budgets~\cite{susskind2022wnnedge}. However, comparatively little attention has been paid to \emph{using WNNs as lightweight attention mechanisms} that \emph{gate} downstream compute in detection pipelines, an attractive role given the extremely low cost of memory lookups at frame rate.

\subsection{Object Detection: From Proposals to Real Time}
Object detection has evolved along complementary lines with distinct efficiency–accuracy trade-offs. Two-stage detectors (e.g., Faster R-CNN) generate region proposals and then refine/classify them, achieving strong recall with modest proposal counts~\cite{ren2015faster}. One-stage families (YOLO, RetinaNet, EfficientDet) predict boxes and classes directly from feature maps to reach real-time throughput with compact backbones~\cite{redmon2016yolo}. Transformer-based detectors (DETR and successors) learn attention end-to-end and simplify post-processing at the cost of higher parameter counts and training compute~\cite{carion2020detr}. 
Across these families, a persistent inefficiency remains: most models expend computation over the full field of view. This contrasts with biological vision and motivates selective processing that concentrates compute where evidence accumulates.

\section{System Overview and Dataflow} 

We propose an end-to-end selective detection pipeline that achieves high accuracy under microcontroller-class power and memory by decoupling \emph{where we look} from \emph{how we detect} (Fig.~\ref{fig:end-to-end}). In \encircle{1}, an always-on DWN runs on the MCU and transforms compact eye/scene inputs into a fixation $\mathbf{p}_t$ and a small set of ROI proposals $\mathcal{R}_t$, with ROI side lengths chosen from $\{48,64,80\}$\,px via the sizing rule $S(p)$. In \encircle{2}, the MCU transmits only lightweight ROI metadata (and optional fixation) to the host over a low-power link, avoiding full-frame transfer early in the pipeline.
On the host, \encircle{3} performs a per-frame spatial union $S_t = \bigcup_i r_t^{(i)}$, and \encircle{4} constructs a single \emph{Union-of-ROIs mosaic} $M_t^{(\mathrm{spatial})} = \mathrm{mosaic}(I_t,S_t)$. The mosaic buffer and detector workspace scale with the processed area $|M_t|$, so the union area directly controls FLOPs, memory traffic, and latency. 
In \encircle{5}, the host maintains a \emph{temporal union} ($U_t = \lambda U_{t-1} \cup S_t$), which accumulates attention over a $K$-frame horizon.
A simple periodic+budget policy governs detector invocations in \encircle{6}: we use the simple setting $R=1$ and run YOLOv12n \cite{tian2025yolov12} on the accumulated mosaic $M_t = \mathrm{mosaic}(I_t, U_t)$ at every frame, while more general $R>1$ policies are supported by the same interface. Detector weights remain fixed while the active compute and memory track $|M_t|$.
Between refreshes, we apply \emph{rotation-/motion-aware ROI stabilization} (yaw/pitch reprojection and IMU-driven inflation) to preserve recall until the next run. Finally, \encircle{7} returns detections (boxes/scores) for application use and optional ROI realignment, keeping high-fidelity digitization and heavy compute confined to the union crop.
The sensor/MCU path supports common modules (e.g., HM01B0/OV2640) and maintains an always-on footprint of only a few KiB; the host performs union/mosaic/temporal fusion and schedules detector refreshes. When $K{=}1$ (no temporal memory), the scheme reduces to the per-frame spatial union.
\begin{figure}[t]
    \centering
    \includegraphics[width=\linewidth]{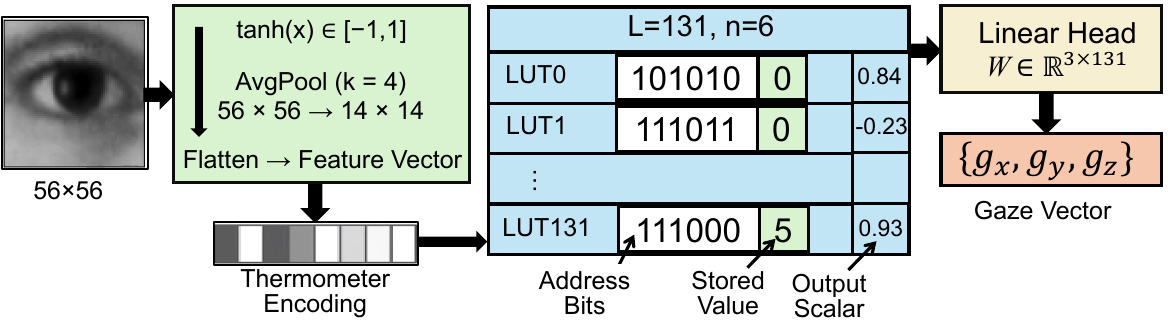}
    \caption{Proposed DWN architecture.} 
    \label{fig:wnn}
    \vspace{-1em}
\end{figure}

\subsection{Weightless Gaze Estimation}
We propose a \emph{DWN} that estimates a 3D gaze direction directly from a normalized grayscale eye image (Fig.~\ref{fig:wnn}).
Each input sample is an eye crop denoted by $x \in \mathbb{R}^{1\times S\times S}$, where $S$ represents the input image size to the DWN. To stabilize illumination and mitigate contrast variations, pixel intensities are first passed through a bounded nonlinearity $\tilde{x}=\tanh(x)$, which compresses outliers while preserving discriminative mid-range gradients around the iris and pupil. The transformed image $\tilde{x}$ is then spatially reduced using parameter-free average pooling with kernel and stride $k$, producing a lower-resolution feature map $u=\mathrm{AvgPool}_k(\tilde{x}) \in \mathbb{R}^{1\times H\times W}$, where $H=W=S/k$. This step preserves the overall geometric structure of the eye while reducing redundancy and memory footprint. Flattening $u$ yields a compact feature vector $\mathbf{f}=\mathrm{vec}(u)\in\mathbb{R}^{F}$, where $F=H\cdot W$ corresponds to the number of pooled spatial features that will later serve as inputs to the discretization stage.
Each scalar feature $f_j$ is discretized using a $K$-bit thermometer encoder that converts continuous values into ordered binary patterns suitable for memory-based table addressing. To adapt the discretization process to feature statistics, per-feature quantile thresholds are computed once on the training data as 
${\small
\tau_{j,k} = \mathrm{Quantile}_{q_k}\!\big(\{f_j^{(n)}\}_{n=1}^N\big),\  q_k=\tfrac{k}{K+1},\; k=1,\ldots,K
}$, ensuring approximately uniform bit activations across features. During training, we maintain differentiability through logistic relaxation
\vspace{-0.2em}
\[
{\small
b^{(\mathrm{soft})}_{j,k}=\sigma\!\left(\frac{f_j-\tau_{j,k}}{T}\right), \qquad \sigma(z)=\tfrac{1}{1+e^{-z}},
}\vspace{-0.2em}
\]
where $T$ controls the temperature of the relaxation. In inference, this soft encoder is replaced by a deterministic hard version $b^{(\mathrm{hard})}_{j,k} = 1 \text{ if } f_j \ge \tau_{j,k}, \; 0 \text{ otherwise}$, producing stable binary codes.
Concatenating all thermometer bits forms a global binary address vector $\mathbf{b}\in\{0,1\}^{B}$, where $B=F\cdot K$ represents the total number of bits and acts as the DWN’s distributed activation pattern.
Computation proceeds through a single layer of small RAM-like LUTs instead of convolution or dense matrix multiplications. The network contains $L$ such tables, each functioning as an associative memory that stores $2^n$ learned entries indexed by $n$-bit addresses. For the $i$-th table, a fixed subset of $n$ bits from $\mathbf{b}$ is selected through a predefined mapping $\Pi_i$, forming a unique binary address that retrieves one of its stored values:
$z_i=\mathrm{LUT}_i\!\big(\Pi_i(\mathbf{b})\big),\  i=1,\ldots,L$.
Stacking all LUT responses yields a distributed latent vector $\mathbf{z}=[z_1,\ldots,z_L]^\top\in\mathbb{R}^{L}$, which encodes the memory-driven representation of the input. During backpropagation, only the entries accessed in each LUT receive gradient updates, resulting in localized, memory-centric learning rather than global weight adjustments.
The LUT responses are aggregated through a lightweight linear projection that maps the distributed memory state into a continuous 3D gaze vector, followed by normalization to ensure unit magnitude:
\vspace{-0.2em}
\[
{\small
\mathbf{y}=\mathbf{W}\mathbf{z}+\mathbf{c}, \qquad \hat{\mathbf{g}}=\frac{\mathbf{y}}{\|\mathbf{y}\|_2}.
}\vspace{-0.2em}
\]
Here, {\small $\mathbf{W}\in\mathbb{R}^{3\times L}$} and $\mathbf{c}\in\mathbb{R}^{3}$ denote the learnable linear head, and $\hat{\mathbf{g}}=[g_x,g_y,g_z]^\top$ represents the predicted 3D gaze direction. The network is trained end-to-end using a composite loss that balances Euclidean and angular alignment with the ground-truth unit vector $\mathbf{g}$; 
{\small $\mathcal{L} = \lambda \|\hat{\mathbf{g}}-\mathbf{g}\|_2^2 + (1-\lambda)\big(1-\langle \hat{\mathbf{g}},\mathbf{g}\rangle\big)$}, where $\lambda\in(0,1)$ adjusts the trade-off between direction and magnitude consistency. Optimization is performed using the Adam optimizer with weight decay and gradient clipping for stability.


\subsection{From Gaze Estimation to ROI Formation}
\begin{figure}[t]
    \centering
    \includegraphics[width=\linewidth]{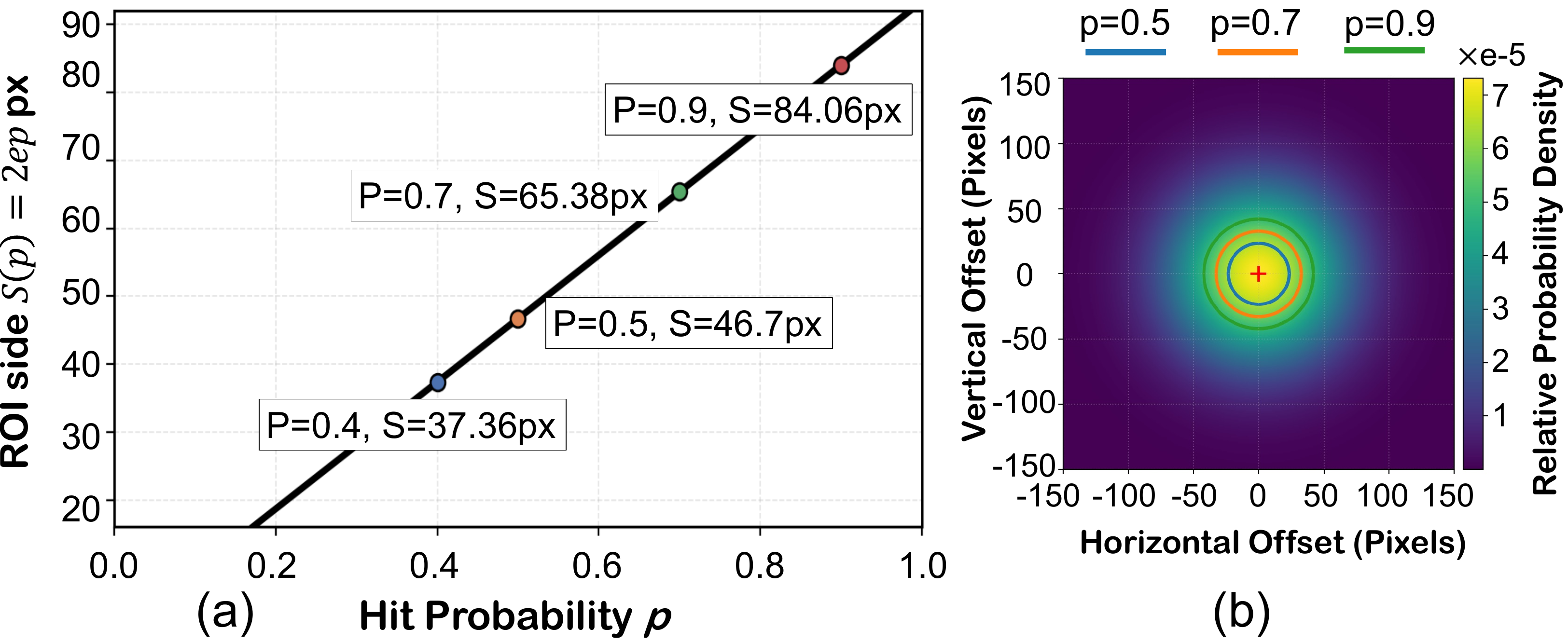}
    \caption{(a) ROI scale from gaze uncertainty. 
    ROI side length as a function of hit probability $p$, with $S(p) ={2ep}$ and $e = 46.7$\,px as the projected uncertainty for an $8.3^{\circ}$ gaze error. (b) ROI selection from gaze uncertainty. 
    Visualization of the 2D gaze uncertainty field ($\mu{=}8.3^{\circ}$, $e{\approx}46.7$\,px) 
    with circular ROIs for
    $p{=}\{0.5,0.7,0.9\}$ mapped to frame sizes of 
    (48$^2$, 64$^2$, 80$^2$\,px).}
    \label{fig:s_of_p}
    \vspace{-1em}
\end{figure}

The normalized gaze $\hat{g}=(g_x,g_y,g_z)$ is projected to the image plane to obtain the fixation $p=(u,v)$. With the horizontal field of view (FOV) and image width $W_{\text{img}}$, an angular uncertainty $\theta_{\max}$ produces a spatial deviation $e$ on the image. This deviation corresponds to how far apart two rays separated by $\theta_{\max}$ appear when they intersect the image plane. The geometric relation between image width and FOV gives $e = \big[W_{\text{img}}/\big(2\,\tan(\mathrm{FOV}/2)\big)\big].\tan(\theta_{\max})$. For $W_{\text{img}}{=}640$ and $\mathrm{FOV}{=}90^\circ$, and with $\theta_{\max}{=}8.3^\circ$, this evaluates to $e \approx 46.7$ px. We define the side length of the region of interest (ROI) as a linear function of the desired hit probability $p\!\in\![0,1]$, $S(p)=2ep$.
Here, $S(p)$ scales proportionally with the confidence level associated with gaze accuracy. 
As illustrated in Fig.~\ref{fig:s_of_p}\,(a), using {\small $e=46.7$}\,px yields ROI side lengths of {\small $S(0.5)=46.7$}\,px, {\small $S(0.7)=65.38$}\,px, and {\small $S(0.9)=84.06$}\,px, which are rounded to {\small $\{48,64,80\}$}\,px for model compatibility and input alignment. 
The corresponding 2D gaze uncertainty distribution and ROI boundaries for $p=\{0.5,0.7,0.9\}$ are visualized in Fig.~\ref{fig:s_of_p}\,(b), representing the spatial distribution from which these scale values are derived.
For each frame $t$, we center a primary ROI at $\mathbf{p}_t$ and (optionally) add a few auxiliary ROIs around high-likelihood neighborhoods (e.g., slight offsets to compensate for micro-saccades). Let the per-frame proposal set be $\mathcal{R}_t=\{r_t^{(i)}\}$ with each $r_t^{(i)}$ a square of side in $\{48,64,80\}$.
We consolidate proposals by a \emph{spatial} union $S_t=\bigcup_{i} r_t^{(i)}$.
 Rather than running the detector on each ROI, we build a \emph{single} image input, the \textbf{\texttt{Union-of-ROIs mosaic}}, from the current frame $I_t$ via $M_t^{(\mathrm{spatial})}=\mathrm{mosaic}(I_t,\tilde{S}_t)$.
Here, $\mathrm{mosaic}(I,\Omega)$ extracts the minimal rectangle covering mask $\Omega$ (tiling if needed) and assembles a single crop, avoiding multiple detector passes and preserving contiguity for objects straddling adjacent ROIs. Pixels belonging to $\tilde{S}_t$ are digitized once and deduplicated, reducing sensor and memory traffic.

\subsection{Attention-Guided ROI-Based Object Detection}
To stabilize recall across motion and partial views, the host maintains a \emph{temporal} union that accumulates per-frame attention over a $K$-frame window:
$U_t=\!\lambda\,U_{t-1}\ \cup\ S_t$,
with optional decay $\lambda\!\in(0,1]$ to down-weight stale regions. The detector input at a refresh event is the 
\textbf{\texttt{accumulated mosaic}}
$
M_t=\mathrm{mosaic}\!\left(I_t,U_t\right)$.
We invoke YOLOv12n under a periodic trigger:
\texttt{run YOLO at} $t$ \texttt{for all} $t$.
After each detector run, $U_t$ will be softly decayed to prevent drift.
Maintaining $U_t$ is linear in the area of incoming ROIs. Detector cost scales with the mosaic area $|M_t|$ rather than the full frame. Consequently, the always-on DWN sustains frame-rate attention at near-zero MACs, while the host amortizes heavy detection over time: \emph{spatial} fusion creates a single Union-of-ROIs mosaic per trigger; \emph{temporal} accumulation preserves recall for moving and partially observed objects. When $K{=}1$ (no memory),  $U_t \;=\; \bigcup_{i=0}^{K-1} S_{t-i}=S_t$ and the temporal scheme reduces to the per-frame spatial union.

To evaluate how \emph{spatial selection alone} affects detection, we use YOLOv12n (nano) as the detector and replace only its input with a gaze-seeded \emph{Union-of-ROIs mosaic} on the COCO validation set. All detector hyperparameters are fixed to match the global baseline (\texttt{conf=0.20}, \texttt{iou\_nms\footnote{Non-Maximum Suppression}=0.70}, \texttt{batch=1}); preprocessing is identical. 
The gaze priors are generated by our DWN, which achieves an average angular error of 8.32$^\circ$ under sub-10\,ms latency. 
Although this accuracy is moderate compared to heavier models, the extremely low latency allows inference at over 100\,Hz, enabling dense temporal sampling that compensates for per-frame angular deviation when forming ROIs across consecutive frames.


\subsubsection{Protocol (spatial, per-frame)}
For each image, an offline, reproducible gaze prior provides a fixation $\mathbf{p}$. We instantiate $N$ fixed-size square ROIs with side $s\in\{48,64,80\}$ centered at $\mathbf{p}$ and (optionally) nearby offsets to model micro-saccades. After geometric overlap control (NMS on ROIs), we form the spatial union $S = \textstyle{\bigcup_{i=1}^{N}r^{(i)}}$, and construct a single mosaic $M=\mathrm{mosaic}(I,S)$. YOLOv12n is run \emph{once} on $M$; all other settings are unchanged. COCO images exhibit substantial variation in native resolution and aspect ratio. 
Following ROI overlay composition, the resulting mosaics are forwarded directly to the YOLO detector without any manual resizing. 
During inference, YOLO performs an internal rescaling of all inputs to its canonical resolution of $640\times 640$, consistent with the official COCO evaluation protocol. 
This automatic normalization ensures that ROI-augmented and baseline images share an identical preprocessing pipeline, thereby isolating the contribution of spatial attention from any confounding effects of image scaling or interpolation. We fix the random seed and evaluate all $(s,N)$ on the same image list for strict comparability.

\subsubsection{Metric}
To isolate cropping effects, we use a single metric, \emph{Accuracy}, with the same hit definition as the YOLO baseline: a GT is correct if any prediction achieves 
$\mathrm{IoU}\ge 0.3$, independent of class. Formally,
\begin{equation}
\mathrm{Acc}(s,N)=\frac{1}{|\mathcal{G}|}\sum_{b\in\mathcal{G}}\mathbf{1}\!\left[\exists\,\hat{b}:\mathrm{IoU}(\hat{b},b)\ge 0.3\right].
\end{equation}

Because the detector architecture and thresholds are fixed, changes in $\mathrm{Acc}$ directly reflect \emph{target recoverability} under restricted visibility. We report results stratified by COCO size (Small: $A<32^2$, MEDium: $32^2\le A<96^2$, LARGE: $A\ge 96^2$).

\subsubsection{Behavior by scale}
For \textbf{small} objects, compact ROIs ($s{=}48$) reduce clutter but may require larger $N$ to compensate for fixation bias. For \textbf{MEDium} objects, $s{=}64$ yields a balanced trade-off, typically saturating around $N\!\in\![20,30]$. For \textbf{LARGE} objects, $s{=}80$ with moderate $N$ restores coverage while maintaining background suppression. These trends indicate that \((s,N)\) jointly govern recoverability and that a \emph{one-pass union mosaic} exposes scale-dependent behavior without altering the detector.

\subsection{Temporal ROI Accumulation (Policy Study)}
Finally, we assess the host-side policy that accumulates ROIs over time. On image sequences (or emulated scanpaths), we maintain the temporal union $U_t=\lambda U_{t-1}\cup S_t$, and invoke YOLOv12n under the periodic trigger:
\texttt{run YOLO at }$t$ \texttt{ for all } $t$.
The detector consumes the \emph{accumulated} mosaic $M_t=\mathrm{mosaic} (I_t,U_t)$. We report  
\begin{equation}
\begin{split}  
\mathrm{Acc}_\mathrm{temp}(s,N,K,R,B),\quad
\overline{|M|}=\frac{1}{T}\sum_t |M_t|,\\
\#\text{refreshes}=\sum_t \mathbf{1}[(t\bmod R{=}0)\ \text{or}\ |U_t|>B],
\end{split}
\end{equation}
capturing accuracy, average processed area (a proxy for compute and energy), and detector invocations. COCO is static; thus, for ablations on COCO we emulate temporal accumulation by ordering proposals into $K$ steps and applying the same trigger logic.
The \emph{spatial} union-of-ROIs mosaic yields a single, detector-ready crop that preserves contiguity while eliminating multiple passes; the \emph{temporal} accumulation amortizes detector cost and maintains recall through simple refresh rules $(R)$. Together, these mechanisms translate DWN's memory-light attention into measurable savings in processed area, energy surrogates, and latency, with accuracy governed primarily by $(s,N)$ and policy $(K,R)$.

\subsection{Rotation- and Motion-Aware ROI Stabilization}
Head motion changes where previously attended regions appear in the image and can enlarge nearby objects between detector refreshes. To preserve spatial coherence and recall at minimal cost, we use two lightweight mechanisms: (i) \emph{head-rotation–aware ROI realignment} into a persistent world view, and (ii) \emph{movement-aware, depth-agnostic ROI inflation} between YOLO runs.

\subsubsection{Head-rotation–aware realignment}
We anchor head orientation to a fixed world reference with yaw–pitch angles $(\psi,\theta)$ and map image coordinates through a pinhole camera. For image width $W$ and horizontal $\mathrm{FOV}$, $f=W/\big(2\tan(\mathrm{FOV}/2)\big)$.
When the pose changes from $(\psi,\theta)$ to $(\psi',\theta')$, the differential rotation $(\Delta\psi,\Delta\theta)$ shifts pixel coordinates as
\[
\begin{bmatrix} u'\\ v' \end{bmatrix}
=\begin{bmatrix} u\\ v \end{bmatrix}
- f \begin{bmatrix} \tan(\Delta\psi)\\ \tan(\Delta\theta) \end{bmatrix},
\]
with the small-angle approximation $\tan \Delta \approx \Delta$ in radians when applicable. We apply this warp to reproject each ROI into a $360^\circ$ \emph{world-view map} (equirectangular panorama) so that successive fixation-driven unions are stitched into a globally consistent layout. 
To avoid recomputation, each union mosaic stored on the host is tagged by quantized $(\psi,\theta)$ and a timestamp for automatic expiration in dynamic scenes. The MCU also keeps a small cache of recent detection boxes with pose tags; upon rotation, it predicts their new positions via the same shift, providing low-latency visual continuity before the next YOLO refresh.


\subsubsection{Movement-aware, depth-agnostic inflation}
Forward motion makes objects appear larger between detector invocations, even though their true depth is unknown. Let $Z_{\min}$ denote a conservative minimum object distance for the environment (smaller indoors, larger outdoors). For frame interval $\Delta t$ and forward acceleration $a_{\mathrm{fwd}}$, the apparent size change satisfies \vspace{-0.25em}
\[
\frac{\Delta s}{s}\;\le\; \frac{|a_{\mathrm{fwd}}|\,\Delta t}{Z_{\min}},\vspace{-0.25em}
\]
so we bound the per-frame isotropic inflation by
$\alpha=\min\!\big(\alpha_{\max},\; \kappa \allowbreak \,|a_{\mathrm{fwd}}|\,\Delta t\big),\  \kappa=1/Z_{\min},$
and update each ROI side as $S' = S(1+\alpha)$ (applied only when $a_{\mathrm{fwd}}>0$; pure rotation does not change apparent scale). Inflated boxes are clamped to image bounds and the crop budget $B$, and persist until (i) new detections overlap them or (ii) a motion-dependent timeout $T$ elapses (faster motion $\Rightarrow$ shorter $T$).
These two operations use only IMU pose/acceleration and simple arithmetic. They add negligible overhead, stabilize the accumulated union $U_t$ under rotation and translation, and preserve recall between YOLO refreshes, while keeping processed area $|M_t|$ (and thus compute/memory) bounded by the budgeted mosaic.
\vspace{-0.5em}

\section{Validation and Evaluation}
\label{sec:eval}


\begin{table*}[t]
  \centering
  \caption{MPIIGaze comparison results: accuracy, computation, and memory footprint.}
   \vspace{-0.5em}
  \begin{adjustbox}{max width=0.99\textwidth}
  \begin{tabular}{lcccccccccc}
  \hline \hline
\multicolumn{1}{c}{\multirow{2}{*}{\textbf{Model}}} & \multirow{2}{*}{\textbf{\begin{tabular}[c]{@{}c@{}}Mean \\ Error (°)\end{tabular}}} & \multirow{2}{*}{\textbf{\begin{tabular}[c]{@{}c@{}}Computation \\ (MACs / LUTs)\end{tabular}}} & \multirow{2}{*}{\textbf{\begin{tabular}[c]{@{}c@{}}Memory\\ (FP32)\end{tabular}}} & \multicolumn{3}{c}{\textbf{Delay(s)}} & \multicolumn{3}{c}{\textbf{Energy(J)}} & \multirow{2}{*}{\textbf{Notes}} \\
\multicolumn{1}{c}{} &  &  &  & \multicolumn{1}{|c}{\textbf{MSP431}} & \textbf{Nano 33 BLE} & \multicolumn{1}{c|}{\textbf{STM32H}} & \textbf{MSP431} & \textbf{Nano 33 BLE} & \multicolumn{1}{c|}{\textbf{STM32H}} &  \\ \hline
    iTracker \cite{krafka2016itracker} & 6.2 & $\sim2.6\times 10^{9}$ / -- & \multicolumn{1}{c|}{108 MiB} & -- & -- & \multicolumn{1}{c|}{--} & -- & -- & \multicolumn{1}{c|}{--} & CNN baseline \\
    SWCNN \cite{zhang2017fullface} & 4.8 & $\sim3.0\times 10^{9}$ / -- & \multicolumn{1}{c|}{11 MiB} & -- & -- & \multicolumn{1}{c|}{8.328} & -- & -- & \multicolumn{1}{c|}{2.748} & Full-face cues \\
    Dilated-Net \cite{chen2018dilatednet} & 4.8 & $\sim1.0\times 10^{9}$ / -- &\multicolumn{1}{c|}{20 MiB} & -- & -- & \multicolumn{1}{c|}{2.776} & -- & -- & \multicolumn{1}{c|}{0.916} & Dilated convs \\
    FAZE \cite{park2019faze} & 5.2 & $\sim3.0\times 10^{9}$ / -- & \multicolumn{1}{c|}{64 MiB} & -- & -- & \multicolumn{1}{c|}{--} & -- & -- & \multicolumn{1}{c|}{--} & Appearance flow \\
    CA-Net \cite{cheng2020canet} & 4.1 & $\sim3.0\times 10^{9}$ / -- & \multicolumn{1}{c|}{82 MiB} & -- & -- & \multicolumn{1}{c|}{--} & -- & -- & \multicolumn{1}{c|}{--} & Attention \\
    SAZE \cite{kim2024saze} & \mytbg 3.89 & $\sim9.0\times 10^{8}$ / -- & \multicolumn{1}{c|}{11.2 MiB} & -- & -- & \multicolumn{1}{c|}{2.498} & -- & -- & \multicolumn{1}{c|}{0.824} & Spherical encoding \\
     \textbf{\texttt{GLANCE}} & \mytbo \textbf{8.32} & \mytbg \textbf{393} / \textbf{131} & \mytbg \textbf{2.2 KiB (8-bit)} & \mytbg \textbf{5.70e-05} & \mytbg \textbf{8.20e-06} & \mytbg \textbf{1.09e-06} & \mytbg \textbf{3.04e-07} & \mytbg \textbf{8.90e-08} & \mytbg \textbf{3.60e-07} & Single-LUT backbone \\ \hline \hline
  \end{tabular}
  \end{adjustbox}
  \label{tab:gaze}
\end{table*}

\begin{figure}[t]
\centering
\includegraphics[width=0.99\linewidth]{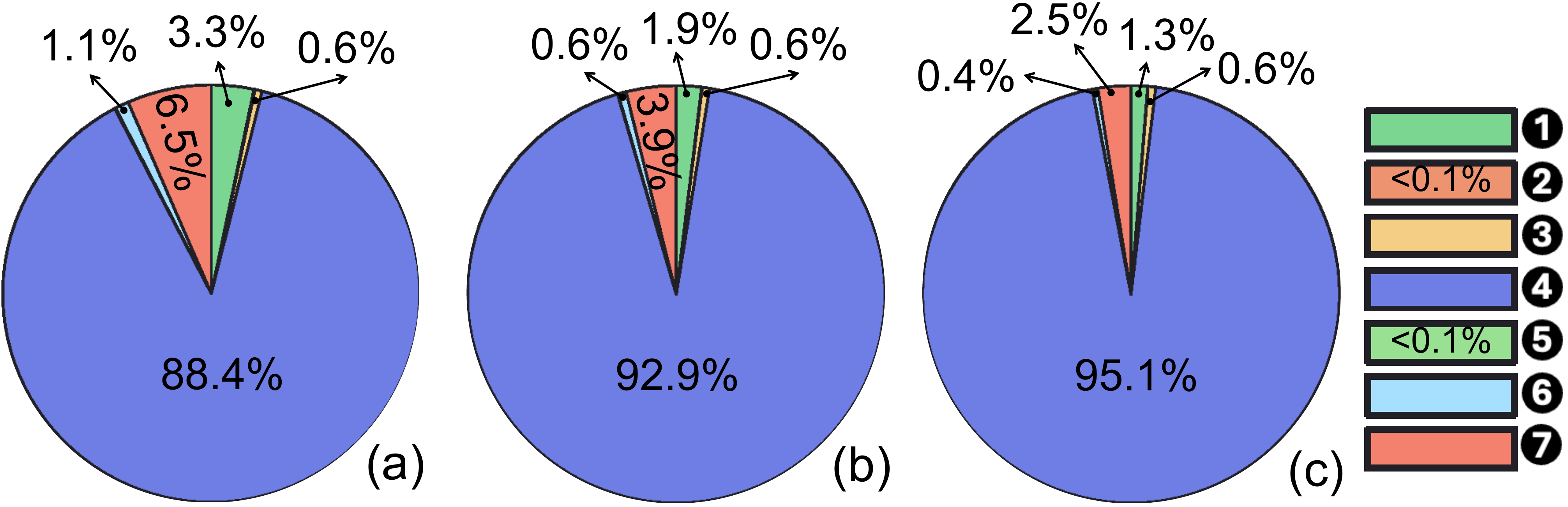}
\caption{\textbf{\texttt{GLANCE}} end-to-end system delay across varying ROI sizes: (a) 48$\times$48, (b) 64$\times$64, and (c) 80$\times$80, measured on the Arduino Nano 33 BLE (microcontroller) and NVIDIA RTX 3090 (host), w.r.t. the required processing steps \protect\encircle{1} -- \protect\encircle{7}.}
\label{fig:roi_delay}
\vspace{-1em}
\end{figure}

\begin{figure}[t]
    \centering
    \includegraphics[width=\linewidth]{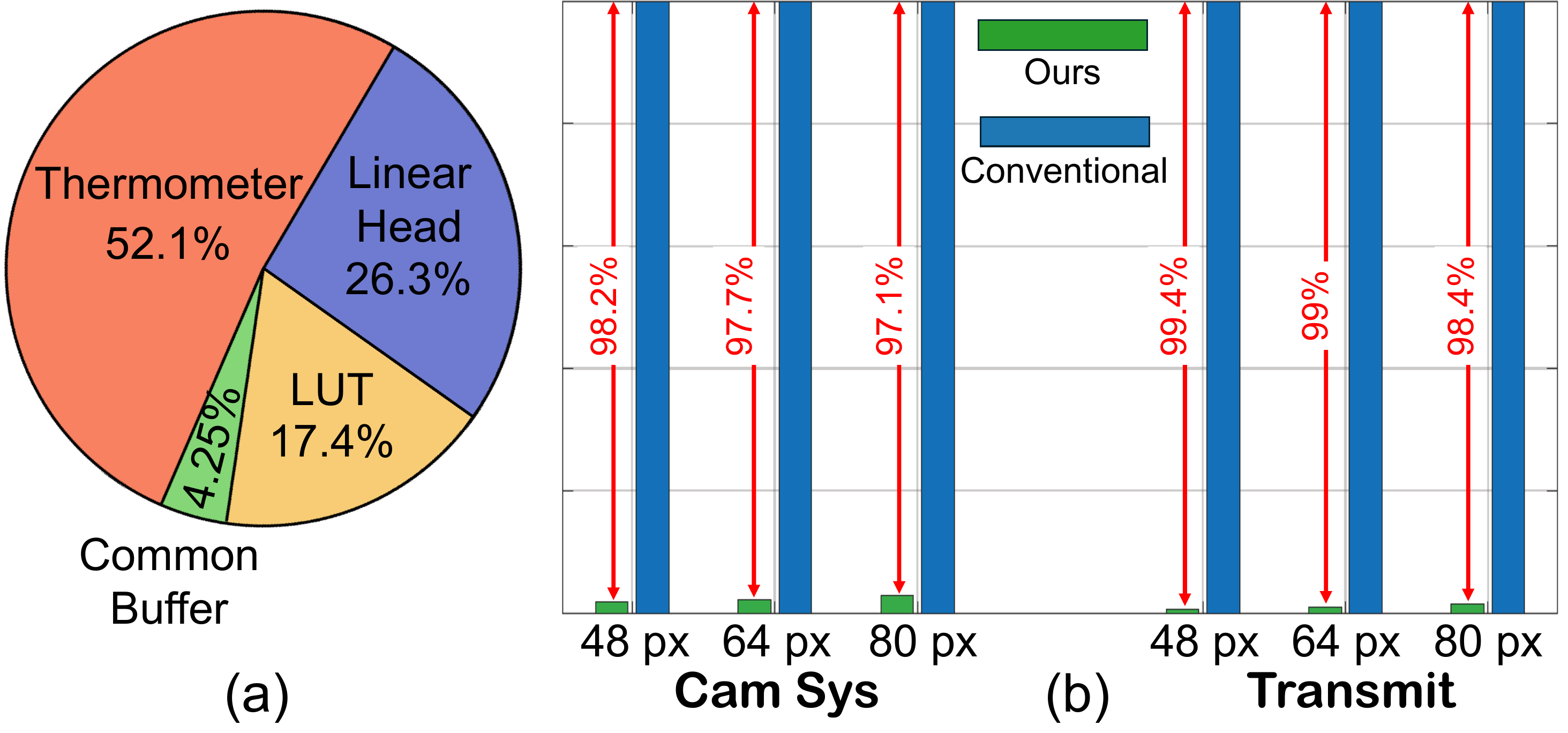}
    \caption{(a) Memory footprint and (b) energy improvement of \textbf{\texttt{GLANCE}} vs. the conventional approach at the MCU end.} 
    \label{fig:comp_engr}
    \vspace{-1em}
\end{figure}

\begin{table*}[t]
\centering
\caption{\textbf{\texttt{GLANCE}} detection accuracy (\%) across ROI counts (K) for different object sizes and ROI patch sizes.}
\vspace{-0.5em}
\label{tab:roi_accuracy}
\resizebox{0.99\textwidth}{!}{
\begin{tabular}{lcccccccccccccc}
\hline \hline
\multirow{2}{*}{\textbf{Object Size}} &
\multirow{2}{*}{\textbf{Count Number}} &
\multicolumn{13}{c}{\textbf{Acc (ROI Count N)}} 
\\
\cmidrule(lr){3-15}
& & \textbf{1} & \textbf{2} & \textbf{3} & \textbf{4} & \textbf{5} & \textbf{10} &
\textbf{15} & \textbf{20} & \textbf{25} & \textbf{30} & \textbf{40} & \textbf{50} 
& \textbf{Global}\\
\midrule
\multirow{3}{*}{\textbf{small}}
& \textbf{48} & 29.55 & 37.96 & \mytbg \textbf{42.67} & 45.62 & 47.94 & 49.06 & 51.34 & 50.72 & 50.24 & 51.26 & 50.67 & 49.25 &  \\
& \textbf{64} & 34.68 & \mytbg \textbf{42.50} & 48.46 & 48.92 & 45.16 & 49.32 & 51.02 & 52.17 & 49.74 & 48.69 & 50.94 & 48.41 & \textbf{42.05}\\
& \textbf{80} & 41.09 & \mytbg \textbf{43.52} & 46.46 & 46.42 & 45.16 & 48.49 & 49.35 & 50.31 & 48.69 & 48.63 & 50.38 & 48.81 & \\
\midrule
\multirow{3}{*}{\textbf{MEDium}}
& \textbf{48} & 53.29 & 59.17 & 61.35 & 64.44 & 66.10 & \mytbg \textbf{72.11}$^{*}$ & 72.79 & 72.67 & 71.99 & 72.84 & 72.66 & 71.99 &  \\
& \textbf{64} & 65.69 & \mytbg \textbf{67.47} & 71.59 & 71.63 & 71.93 & 71.82 & 71.62 & 72.31 & 72.89 & 72.14 & 72.11 & 71.90 & \textbf{67.16}\\
& \textbf{80} & 65.03 & \mytbg \textbf{69.20} & 70.03 & 69.78 & 69.94 & 70.30 & 71.31 & 70.86 & 71.56 & 71.32 & 71.58 & 71.56 & \\
\midrule
\multirow{3}{*}{\textbf{LARGE}}
& \textbf{48} & 23.67 & 48.45 & 60.34 & 64.10 & 68.80 & 80.44 & 84.62 & 86.63 & 87.49 & \mytbg \textbf{88.10}$^{\dag}$ & 89.16 & 89.85 &  \\
& \textbf{64} & 37.05 & 60.15 & 70.43 & 73.03 & 77.37 & 84.82 & 80.33 & \mytbg \textbf{88.36}$^{\diamond}$ & 89.22 & 89.22 & 89.95 & 90.40 & \textbf{87.55}\\
& \textbf{80} & 47.00 & 66.72 & 75.54 & 77.31 & 80.92 & 86.34 & \mytbg \textbf{88.23}$^{\circledast}$ & 89.28 & 89.65 & 90.55 & 90.37 & 90.77 & \\ \hline \hline
\end{tabular}}
\begin{flushleft}
\footnotesize{First $N$ (ROI counts) for which accuracy exceeds the global baseline:
$^*$ROI= 6 $\rightarrow$ 68.32\%. {$^\dag$}ROI= 26 $\rightarrow$ 87.76\%. {$^\diamond$}ROI= 18 $\rightarrow$ 87.85\%. ${^\circledast}$ROI= 13 $\rightarrow$ 87.63\%.}
\end{flushleft}
\vspace{-1.25em}
\end{table*}
To evaluate our framework, we emulate the complete processing pipeline using an Arduino Nano 33 BLE as the edge device and an NVIDIA RTX 3090 GPU as the host. The delays for all processing steps in Fig.~\ref{fig:end-to-end} are shown in Fig.~\ref{fig:roi_delay}. As seen in the figure, most of the time is spent on communication between \textbf{\texttt{GLANCE}} and the host, highlighting the importance of performing preprocessing on the edge device. Instead of transmitting the full 640$\times$640 image, \textbf{\texttt{GLANCE}} sends only small ROI crops (48–80\,px), reducing communication time by approximately $177\times$.
As mentioned, this preprocessing is performed by the proposed DWN architecture, which requires an additional camera to estimate the gaze location. The memory footprint of the network is shown in Fig.~\ref{fig:comp_engr}(a), which is explained in detail in the following section.  
To have a fair comparison for the additional camera, the energy consumption of the gaze camera is also plotted in Fig.~\ref{fig:comp_engr}(b). In this plot, for our setup, 'Cam Sys' includes two low-resolution cameras (up to 80$\times$80 pixels) along with their minimal processing, whereas the conventional system utilizes a single 640$\times$640 camera.
Additionally, in the transmission stage, we assume that \textbf{\texttt{GLANCE}} only needs to transfer the ROI, whereas the conventional system must transmit the entire image each time.
Based on the figure and Table~\ref{tab:roi_accuracy}, 
even in the worst case, detecting large objects with 48\,px ROIs that require 26 frames to match the global baseline, \textbf{\texttt{GLANCE}} still reduces the amortized processed area and thus communication energy, by $\approx$ 85.37\% and $\approx$ 69.3\%, respectively, compared to transmitting the full frame.

\subsection{Gaze Estimation on MPIIGaze}
We evaluate the DWN gaze estimator on \textbf{MPIIGaze}~\cite{zhang2015mpiigaze} using the official normalized eye images and report the mean angular error (degrees). Unless otherwise noted, we follow the standard \emph{Leave-One-Person-Out (LOPO)} protocol: for each held-out subject \texttt{pXX}, all remaining participants are used for training; model selection for that subject uses its validation split. This measures cross-person generalization, which is critical for deployment to unseen users.
To preserve gaze geometry, we do not apply photometric or geometric augmentations. Each image is converted to grayscale, resized to $S\times S$ (LANCZOS), and normalized to $[-1,1]$. Ground-truth gaze vectors $\mathbf{g}=[g_x,g_y,g_z]^\top$ are normalized to unit length,
\[
\hat{\mathbf{g}}=\frac{\mathbf{g}}{\|\mathbf{g}\|_2+\epsilon},\qquad \epsilon=10^{-8},
\]
and samples with near-zero magnitude are replaced by the forward direction $[0,0,1]^\top$ for numerical stability.

\subsubsection{Configuration}
We use $S{=}56$, pooling $k{=}4$ resulting in $F{=}196$ features. The thermometer encoder uses $K{=}4$ bits per feature with training temperature $T{=}0.5$ (hard encoder at inference). The DWN backbone has a single layer with $L{=}131$ LUTs, address width $n{=}6$, followed by a linear head. Training uses Adam (lr $3{\times}10^{-3}$), weight decay $10^{-5}$, batch size $64$, gradient clipping $\|\nabla\|_2\le 5.0$, and the hybrid objective
{\small $\mathcal{L}=\lambda\|\hat{\mathbf{g}}-\mathbf{g}\|_2^2+(1-\lambda)\big(1-\langle \hat{\mathbf{g}},\mathbf{g}\rangle\big)$}, with $\lambda{=}0.3$.

\begin{figure}[b]
\vspace{-1em}
    \centering
    \includegraphics[width=\linewidth]{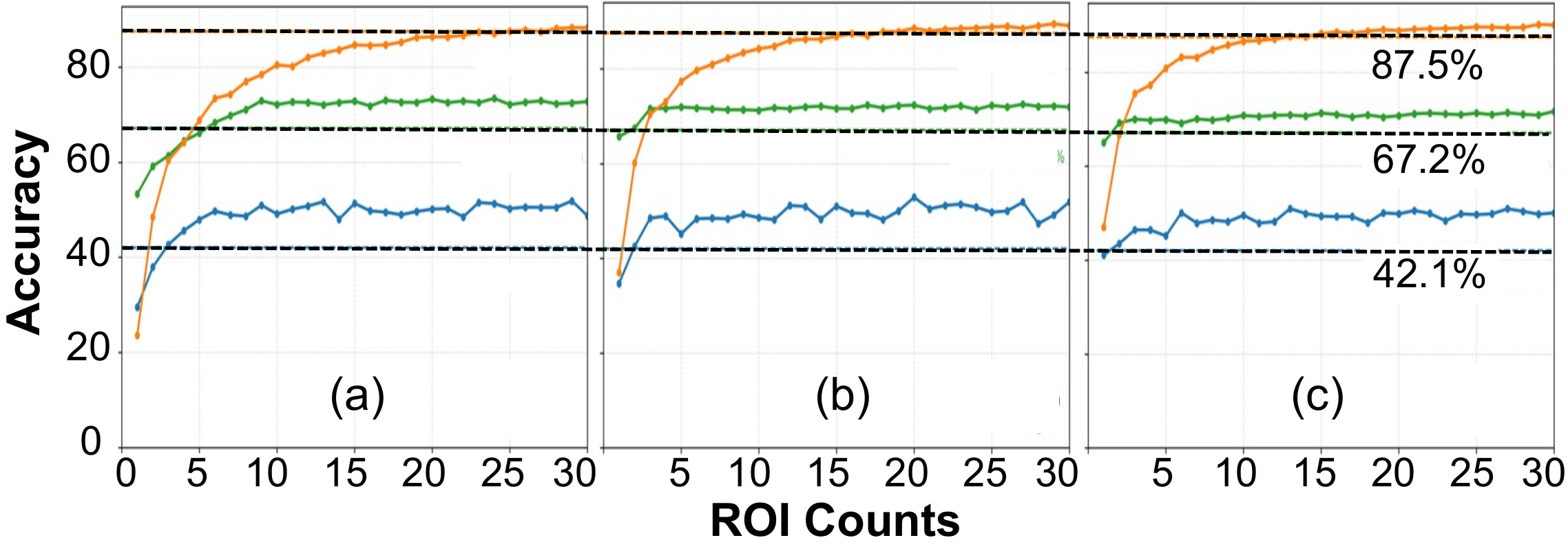}
    \caption{Comparative analysis of ROI-based detection accuracy under increasing ROI counts (up to 30) for ROI sizes of (a) 48 px, (b) 64 px, and (c) 80 px, illustrating performance convergence for different object sizes.}
    \label{fig:roi_convergence}
\end{figure}

\subsubsection{Complexity accounting}
The total parameter count is
$\mathrm{Params} = FK + L2^{n} + (3L + 3),$
corresponding to thermometer thresholds, LUT entries, and the linear head. With $F{=}196$, $K{=}4$, $L{=}131$, and $n{=}6$, this yields $\sim$9.6K parameters. In our MCU-oriented implementation, thermometer thresholds and the linear head are stored in 8-bit precision and LUT entries in 1-bit form, resulting in a parameter footprint of $\sim$2.2\,KiB. The exported PyTorch model artifact used for deployment occupies 0.036\,MB, which includes these parameters together with metadata and library bookkeeping. Inference requires only 131 table lookups and 393 MACs per image, and under this configuration our DWN achieves a mean angular error of \textbf{8.32$^\circ$} on MPIIGaze (subject \texttt{p00}), as summarized in Table~\ref{tab:gaze}.

To validate the performance of our method, we simulated it on three different microcontrollers: the MSP430FR5969 at 16~MHz, the Arduino Nano 33 BLE at 64~MHz, and the STM32H743VIT6 at 480~MHz. As shown in Table~\ref{tab:gaze}, none of the models could run on the first two microcontrollers due to their large parameter requirements, indicated by \textbf{\texttt{--}}, which means ``not applicable'' in the table. For the STM32H743VIT6, 64~MB of external flash memory is required to fit the model. In this scenario, our DWN achieves up to a $2.3\times10^{6}$ reduction in computational cost compared to prior models, within MCU-class device memory limits.
Unlike prior gaze estimators in Table~\ref{tab:gaze}, whose 10–100\,MB footprints require off-chip DRAM or are infeasible on MCUs, our $\sim$9.6K-parameter DWN fits entirely within the on-chip memory of MSP430, Arduino Nano 33 BLE, and STM32 devices, enabling truly always-on gaze estimation without external flash or accelerators.

\subsection{ROI Detection Accuracy and Efficiency Frontier}
The quantitative trends in Table~\ref{tab:roi_accuracy} and the convergence curves in Fig.~\ref{fig:roi_convergence} illustrate how detection accuracy evolves with increasing ROI count~$N$ under constrained spatial budgets. Accuracy rises rapidly in the early stages, indicating that a small number of high-confidence regions already encode the principal task-relevant cues. As $N$ continues to grow, the gain gradually saturates, revealing an \emph{information-completion boundary} where additional regions contribute contextual redundancy rather than novel evidence. This boundary reflects the transition from sparse discriminative reasoning to spatially exhaustive processing, defining the natural limit of efficiency in region-based perception.
A closer examination across object scales uncovers a distinctive \emph{inverse-scale convergence pattern}. Small objects first surpass the global baseline (42.05\%), followed by MEDium objects (67.16\%), while LARGE objects converge last toward their global ceiling (87.55\%). This reversal indicates that ROI-based inference compensates most effectively in the regime where global detectors are weakest. For small targets, which occupy a few pixels and suffer from low contrast, global convolutional aggregation disperses discriminative gradients, resulting in spatial underutilization. In contrast, the ROI mechanism concentrates representational capacity on compact high-saliency regions, producing early and pronounced accuracy gains. As object size increases, the advantage diminishes because large objects are already well-captured by the global receptive field; hence, ROI refinement yields smaller marginal benefits.
These results suggest that ROI-based attention functions as an \emph{adaptive resolution allocator}: it amplifies weak signal zones rather than uniformly redistributing resources. The effect is particularly visible in the sublinear-to-superlinear transition of small-object accuracy curves, where local focus transforms information sparsity into measurable gain. This emergent property demonstrates that selective inference can surpass global completeness locally, not by expanding representation, but by optimizing the \textit{information-to-area ratio}.
ROI patch size further regulates this behavior. Enlarging each patch from 48\,px to 80\,px yields consistent improvements across all scales, implying that broader contextual integration enhances reasoning without diffusing attention. The smoother saturation observed at 80\,px demonstrates more balanced spatial aggregation and reduced volatility in the accuracy frontier. At this setting, the ROI framework approaches the global upper bounds across all scales while maintaining a dramatically lower spatial budget.

\section{Conclusion}
\textbf{\texttt{GLANCE}} shows that memory-centric computation with explicit attention can outperform uniform full-frame processing for edge vision. A single-layer DWN gaze estimator (8.32$^\circ$ error on MPIIGaze with $\sim$9.6K parameters, 131 LUTs, and 393 MACs) drives a union-of-ROIs detection pipeline that achieves 51.3\%/72.1\%/88.1\% accuracy on small/MEDium/LARGE objects, exceeding the 39.2\%/63.4\%/83.1\% global YOLOv12n baseline while processing only 40–50\% of the image area. \textbf{\texttt{GLANCE}} reduces communication latency by 177$\times$ and energy by 65\%, yet maintains end-to-end latency of sub-10\,ms. 
These results establish WNN as a practical ultra-low-power attention front-end for gating modern detectors on severely resource-constrained wearable and embedded vision platforms.

\bibliographystyle{IEEEtran}
\bibliography{IEEEabrv,./Ref}\vspace{-2em}

\end{document}